\documentclass{WileyMSP-template}

\usepackage[utf8]{inputenc}
\usepackage[T1]{fontenc}
\usepackage{textcomp}
\usepackage{xcolor}
\usepackage{graphicx}
\usepackage{dcolumn}   
\usepackage{multicol}
\usepackage{amsmath, mathptmx, mathtools}
\usepackage{caption, subcaption}
\usepackage{float}

\begin{document}

\pagestyle{fancy}
\rhead{\includegraphics[width=2.5cm]{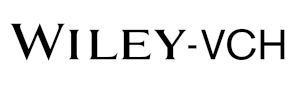}}

\title{Gain-controlled Broadband Tuneability in Mode-locked Thulium-doped Fibre Laser through Variable Feedback}

\maketitle


\author{Dennis C. Kirsch*}
\author{Anastasia Bednyakova}
\author{Benoit Cadier}
\author{Thierry Robin}
\author{Maria Chernysheva}


\begin{affiliations}
D. C. Kirsch\\
Ultrafast Fibre Laser, Leibniz Institute of Photonic Technology, Albert-Einstein-Str.~9, 07702~Jena, Germany, 
Email Address: dennis.kirsch@leibniz-ipht.de

A. Bednyakova\\
Novosibirsk State University, Novosibirsk, 630090, Russia,
Email Address: a.bedniakova7@g.nsu.ru

Benoit Cadier\\
iXblue Photonics, Rue Paul Sabatier 22300 Lannion, France
Email Address: benoit.cadier@ixblue.com

Thierry Robin\\
iXblue Photonics, Rue Paul Sabatier 22300 Lannion, France

M. Chernysheva\\
Ultrafast Fibre Laser, Leibniz Institute of Photonic Technology, Albert-Einstein-Str.~9, 07702~Jena, Germany, 
Email Address: Maria.Chernysheva@leibniz-ipht.de

\end{affiliations}




\begin{abstract}

Broadband wavelength tuneability can ensure a new level of versatility for laser systems and extend areas of their applications. Principle limitations of achieving wide tuning wavelength ranges are generally defined by the spectral bandwidth of the gain and traditional tuneability techniques, relying on electrically controlled or bulk filters, reducing laser efficiency and generation stability. In this work, we present nearly 90~nm tuneability in ultrafast Tm-doped fibre laser within the span from 1873 to 1962~nm by implementing variable feedback for efficient control of the excitation level of the active medium and, hence, the gain spectrum. The highest laser efficiency is observed with 20\% feedback, generating 580-fs soliton pulses at 1877~nm central wavelength with 1.5~nJ output pulse energy. By combining the nonlinear Shrödinger equation and population inversion rate equations for the gain medium, the developed numerical model helps to unveil nonlinear pulse evolution under the influence of dynamically varying gain spectrum. The resulting laser system presents a compact and straightforward approach to achieve laser generation with a broad range tuneability of wavelength and operation regimes, which is not impaired by the limitations on laser stability or power performance and can be translated to other wavelength ranges.
\end{abstract}


\section{Introduction}

Strengthening positions of laser technologies in scientific, industrial and daily life applications relies essentially on the laser system performance and how flexible they can adapt to the broad range of requirements. Versatile tuneable or reconfigurable photonics instruments are, therefore, currently in high demand for developing new empowering technological platforms and making them more sustainable.

In this perspective, Thulium doped fibres enable generation not just at a very desirable wavelength range around 2~$\mu$m, which places these fibre laser systems at the forefront for diverse applications. Environmental monitoring of greenhouse gases~\cite{Moro.2011w, Singh.2017f}, polymer or semiconductor machining~\cite{Acherjee.2020l,Astrauskas.2021i}, optical coherence tomography~\cite{Ota:20211.8,Liang.2013o}, nonlinear microscopy~\cite{Xu.2013r, Nomura.2020s}, and optical communication~\cite{Gunning.2019t, Lin.20202}, these are just a handful of the laser tasks, which were enabled or enhanced with the development of new Tm-doped fibre laser systems. More importantly, Tm-doped fibres feature a remarkable broad gain spectrum (almost 350~nm), allowing to tailor operational wavelength to a particular application. With the use of wavelength selection mechanisms, a tuning range up to 300~nm (1733--2033~nm) has been experimentally demonstrated in ultrafast Thulium-doped fibre lasers~\cite{Dai.2019n}.

However, a broad range of wavelength tuneability is generally realised using special laser components. Typically, these are bulk elements, such as acousto-optic tunable filters, diffraction gratings, volume Bragg gratings, or thin-film coatings, which benefits come at the cost of generation stability and substantial insertion losses. Thus, the possibilities of shaping and manipulating the laser generation through the interplay of intrinsic intracavity phenomena have recently become a topic of intensive experimental and theoretical investigation~\cite{sidorenko2019nonlinear,perego2018gain,tarasov2016mode}.



A full of promise technique to achieve a broadband laser tuneability,  omitting the application of intracavity filters, relies on manipulating gain excitation level. Specifically, in a quasi-three-level model, the equilibrium between emission and absorption, and, thus, the profile of the gain spectrum $g(\lambda)$ is governed by the lower $N_1$ and excited laser level $N_2$ population fractions\cite{Pask.95y, Franco.1994c}:

\begin{equation} 
g (\lambda) = \int_{0}^{z=\ell} dz \{N_2 (z) \sigma_{21} (\lambda) \Gamma - N_1 (z) \sigma_{12} (\lambda) \Gamma\},
\label{alpha}
\end{equation}

\begin{figure}[!t]
 \centering
  \includegraphics[width=0.48\linewidth]{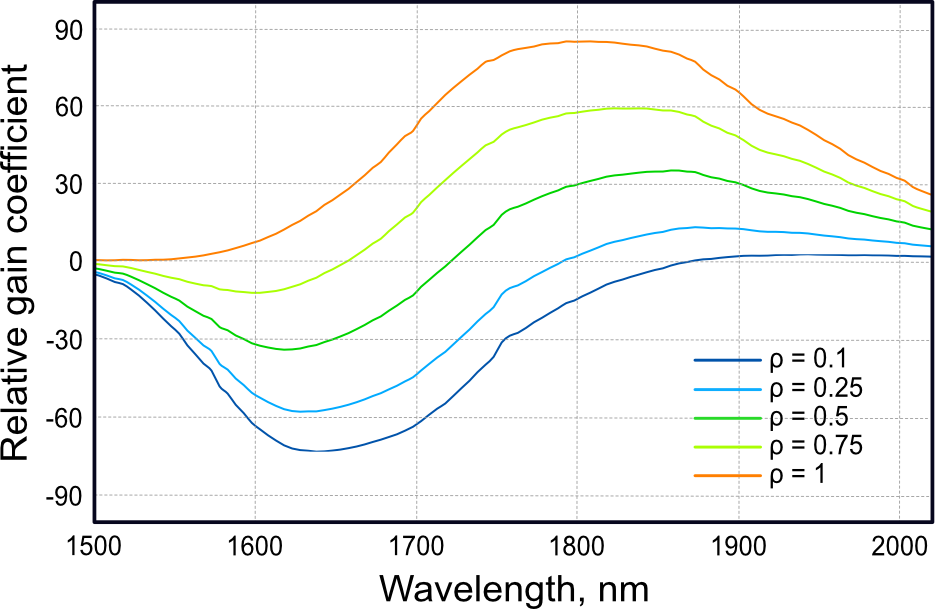}
  \caption{Calculated Tm-doped fibre gain spectra at different energy levels population density.}
  \label{gain}
\end{figure}

\noindent{}here $\ell$ -- is the length of the active fibre, $\sigma_{21}$ and $\sigma_{12}$ -- are emission and absorption cross-sections, correspondingly, and ~$\Gamma$ -- is the confinement factor between dopant and mode field area, respectively. Quantitatively the excitation level can be considered using a ratio $\rho$ of population at $N_2 = N\cdot\rho$ and $N_1= N\cdot(1-\rho)$ levels, where $N$ -- is a total number of all rare-earth ions integrated over the fiber mode cross-section. This model applies to Tm-doped fibres with $^3$H$_6 \rightarrow ^3$F$_4$ pump scheme. \textbf{Figure~\ref{gain}} predicts spectral gain profiles of Tm-doped fibre (used in further investigations) to red-shift at higher excitation level\;\cite{Engelbrecht.08w, ixblueFibers, Chen.2019u, Agger.2006e}.

Control of the laser cavity Q-factor allows efficient governing of an excitation level of gain, leveraging output power and laser generation wavelength. The control of the Q-factor has been realised, for example, by changing the active fibre length or its doping concentration~\cite{Franco.1994c}, or by altering intracavity losses~\cite{dong2005output,Lin.2006L}. Yet, there are no active means of manipulating the parameters of active fibre during laser operation, but only at the laser development stage. At the same time, the intracavity loss modulation has been applied in several research works for filter-less tuneability, typically by implementing variable attenuators \cite{Ghosh2016a,Anjali.2016d,Sharma.2012d,Deepa.2007l,Franco.1994c}. For instance, theoretical work predicted a 105~nm tuneability range for continuous-wave Tm-doped fibre laser~\cite{Sharma.2012d}. However, only 36\,nm-long tuneability has been verified empirically~\cite{Ghosh2016a}. Remarkably, bending of the active fibre presents another technique for easy variation of the operation wavelength and, particularly, establishing generation at shorter ranges, e.g. around 1700~nm in Tm-doped fibre lasers. Thus, 152~nm wavelength tuneability from 1740 to 1892~nm has been demonstrated in normal dispersion Tm-doped ultrafast fibre laser~\cite{chen2020all}. However, the fundamental principle of such tuneability does not rely on the variation of stored energy in the cavity due to increased losses. Instead, bending introduces wavelength-selective losses, suppressing amplified spontaneous emission at the longer edge of the gain spectrum \cite{foroni2006s}. Importantly, like for any other technique relying on loss introduction, these results came at the price of reduced laser efficiency. With up to 1.3~W pump power, the average output power reached only 4.5~mW at maximum.

This work demonstrates effective continuous wavelength tuneability in Tm-doped mode-locked fibre laser by controlling the cavity feedback and maintaining high quantum efficiency in ultrafast Tm-doped fibre laser. We employ a variable output coupler, which determines the intracavity energy and, therefore, the excitation level of Tm-doped fibre gain. To reduce the laser complexity and the influence of other components on the tuneability range, we design a self-mode-locked laser by using active Tm-doped fibre both as a gain medium and saturable absorber. Nevertheless, the complete analysis of the self-mode-locking generation regime is outside the scope of the current work and will be presented in the follow-up study. We experimentally demonstrate the stable generation of femtosecond soliton pulses tuneable within the range spanning from 1873 to 1962~nm with the maximum average output power reaching 83~mW. Our experimental observations are confirmed with the results of numerical simulations, which allowed us to clarify the origin of the tuneability effects and highlighted the pivotal role of optimisation of the saturation and gain level, and glass matrix composition in the defining the tuneability range. The results of our studies show the high potential of laser feedback variation to serve as an efficient mechanism of broadband tuneability. The advantage of the demonstrated technique is that it can be translated to other laser configurations and wavelength ranges, which currently lack an extensive selection of components, such as Mid-IR, to develop robust fibre-based systems.


\section{Results}

\subsection{Experimental setup}

\begin{figure}[!bp]
\centering
  \includegraphics[width=0.45\linewidth]{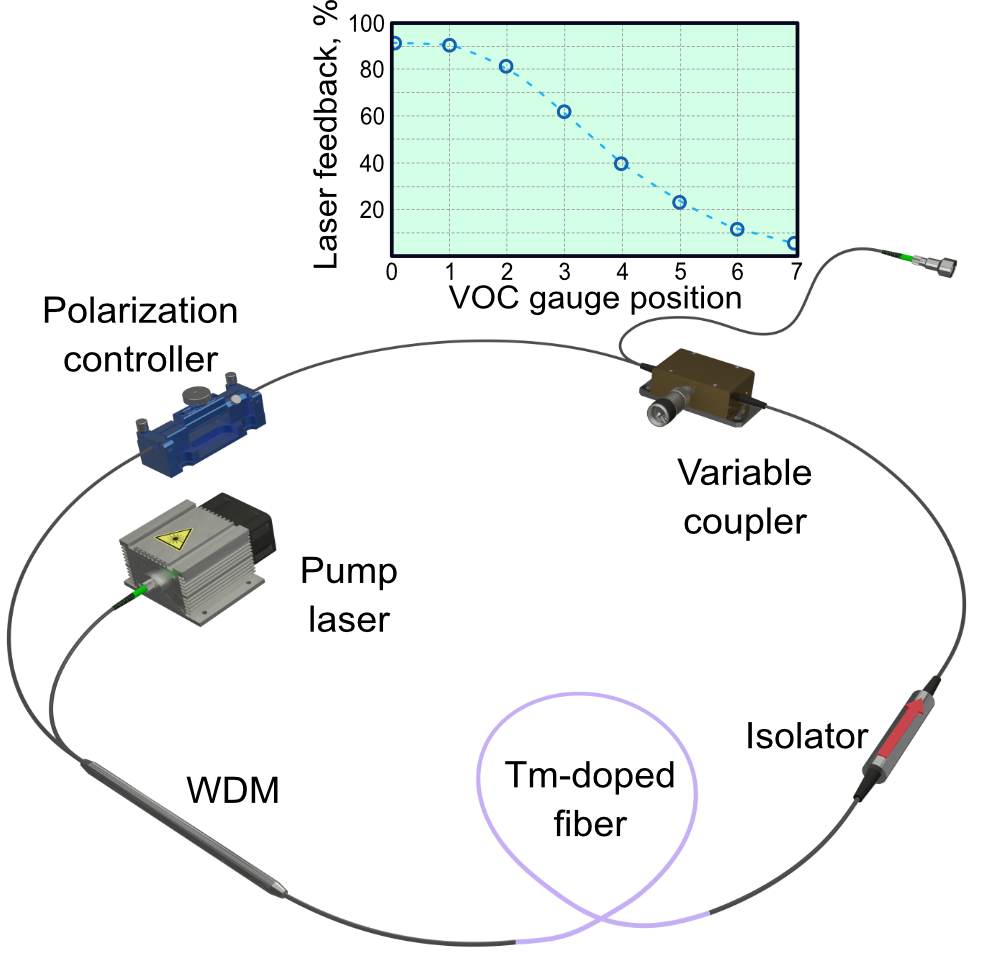}
  \caption{Schematic of the ultrafast Tm-doped fibre laser. Inset: variation of coupling coefficient.}
  \label{schematic}
\end{figure}

To examine the tuneability dynamics of ultrashort pulse generation in the laser cavity with variable Q-factor, we assembled the Tm-doped fibre laser setup as presented schematically in \textbf{Figure~\ref{schematic}}. Ring fibre laser cavity employed 1550-nm pump laser with the power up to 1~W (HPFL-300, BKtel), 1550/2000 wavelength division multiplexer, isotropic isolator and squeezing polarisation controller. A 0.5~m section of a Tm-doped silica fibre (iXblue Photonics) provided both active gain and saturable absorption to enable self-mode-locking. The saturable absorption behaviour originates from $^3H_4$, $^3H_4$~$\rightarrow$~$^3H_6$, $^3F_4$ energy transfer upconversion process~\cite{jackson.2002d}, which is facilitated by formation of ion-pair clusters due to high doping concentration~\cite{sanchez1993effects,Tao.2018i}. As noticed above, the investigation of the self-mode-locking phenomenon in the Tm-doped fibre would be presented in the follow-up paper. The active fibre core was 2.65~\textmu{m} in radius with the numerical aperture of 0.17 and featured $\mathrm{1.8\cdotp10^{26}\,m^{-3}}$ doping concentration. The fibre small-signal gain was experimentally estimated as $\sim$30\,dB (for 0.95\,W pumping and 65\,µW seeding a 44\,cm piece). The active fibre group velocity dispersion,  third-order dispersion and nonlinearity were estimated as $\beta_2$= -20~ps$^2$/km, $\beta_3$ = 0.27~ps$^3$/km, $\gamma$ = 2~W$^{-1}$km$^{-1}$, correspondingly. The rest of the cavity length comprises 4.2 meters-long single-mode fibre ports of fiberised laser components. The passive fibre was non-polarisation maintaining and had following parameters: $\beta_2$~=~-59~ps$^2$/km, $\beta_3$~=~0.28~ps$^3$/km, $\gamma$~=~1.3~W$^{-1}$km$^{-1}$.

The tuneability of the laser generation was established introducing no filtering elements into the cavity, but only through including an in-line variable fibre-optic coupler (Evanescent Optics) with a coupling ratio ranging from 8 to 93\% and 0.3--0.1\,dB excess loss. The variable coupler is based on D-shaped polished fibres, interacting via evanescence field. The separation of these fibres changes through rotating the knob controller, thereby modifying the coupling efficiency from one waveguide into another and overall laser cavity feedback. The performance of the variable coupler was characterised by reference laser transmission measurements at 1.95\,$\mu$m and is shown in the inset in Figure~\ref{schematic}. Such laser configuration is highly integrated. Due to a few required components, it is cost-effective and easy for experimental implementation, yet features rich dynamics of the laser radiation tuneability.

\subsection{Numerical simulations}

To validate the possibility of the broadband tuneability in the mode-locked Tm-doped fiber laser discussed above, we developed first a numerical model, which describes consequent pulse propagation through different cavity elements. 
Pulse propagation along the passive fibre is governed by the Nonlinear Schr$\mathrm{\ddot{o}}$dinger equation \cite{AgrawalNFO4}. The response of saturable absorber on instantaneous pulse power within each round-trip is described with a standard model: $\alpha = \alpha_0/(1+\frac{P}{P_{sat}})+\alpha_{ns}$, where $\alpha$ -- is the intensity-dependent absorption coefficient. In the model, we assumed the linear limits of saturable absorption and non-saturable absorption of $\alpha_0$ = 0.15 and $\alpha_{ns}$ = 0.85, respectively, and saturation power (the power necessary to reduce the absorption coefficient to half the initial value) $P_{sat}$ =100\,W. The system of coupled equations for continuous-wave pump and pulsed signal generation, taking into account the effects of dispersion and nonlinearity, was considered to describe amplification of the signal :

\begin{equation}
\frac{\partial A_{s}(z,t)}{\partial z} =  -i\frac{\beta_2}{2} \frac{\partial^2 A_s(z,t)}{\partial t^2} + i\gamma|A_s(z,t)|^2 A_s(z,t) + \frac{g_s(\omega,z)}{2}A_s(z,t), 
\end{equation}
\begin{equation}
\frac{\partial P_{p}(z)}{\partial z} = g_p(z)P_p(z),
\end{equation}

where $A_s(z,t)$ is the slowly varying envelope associated with signal, $P_p(z)$ is the average power of continuous-wave pump, $\beta_2$ is the group velocity dispersion, $\gamma$ is the Kerr nonlinearity, $g_s$ and $g_p$ are signal and pump gain/loss coefficients. Gain/loss coefficients were found by solving rate equations in the stationary case $dN_2/dt = 0$ \cite{Turitsyn:11,Chen:12}:

\begin{equation}
\frac{dP_s(\lambda_i,z)}{dz} = g_s(\lambda_i,z)P_s(\lambda_i,z) = \left( \sigma_{21}^s(\lambda_i) + \sigma_{12}^s(\lambda_i)\right)\rho_s(\lambda_i)P_s(\lambda_i,z)N_2(z)-\sigma_{12}^s(\lambda_i)\rho_s(\lambda_i)NP_s(\lambda_i,z)
\end{equation}

\begin{equation}
\frac{dP_p(z)}{dz} = g_p(z) P_p(z) = \left( \sigma_{21}^p + \sigma_{12}^p\right)\rho_pP_p(z)N_2(z)-\sigma_{12}^p\rho_p(0)NP_p(z),
\end{equation}

\begin{align}
\frac{dN_2(z)}{dt}=\left( \sigma_{12}^p \rho_p\frac{P_p(z)}{h\nu_p} + \sum_{k=1}^{k} \left( \sigma_{12}^s(\lambda_k) \rho_s(\lambda_k)\frac{P_s(\lambda_k,z)}{h\nu_k}\right) \right)N- \\ \nonumber
\left( \left( \sigma_{21}^p + \sigma_{12}^p\right) \rho_p\frac{P_p(z)}{h\nu_p} + \sum_{k=1}^{k} \left( \left( \sigma_{21}^s(\lambda_k) + \sigma_{12}^s(\lambda_k)\right) \rho_s(\lambda_k)\frac{P_s(\lambda_k,z)}{h\nu_k}\right) + \frac{1}{T}\right)N_2(z),
\end{align}

\begin{figure}[!]
    \centering
    \includegraphics[width=0.95\textwidth]{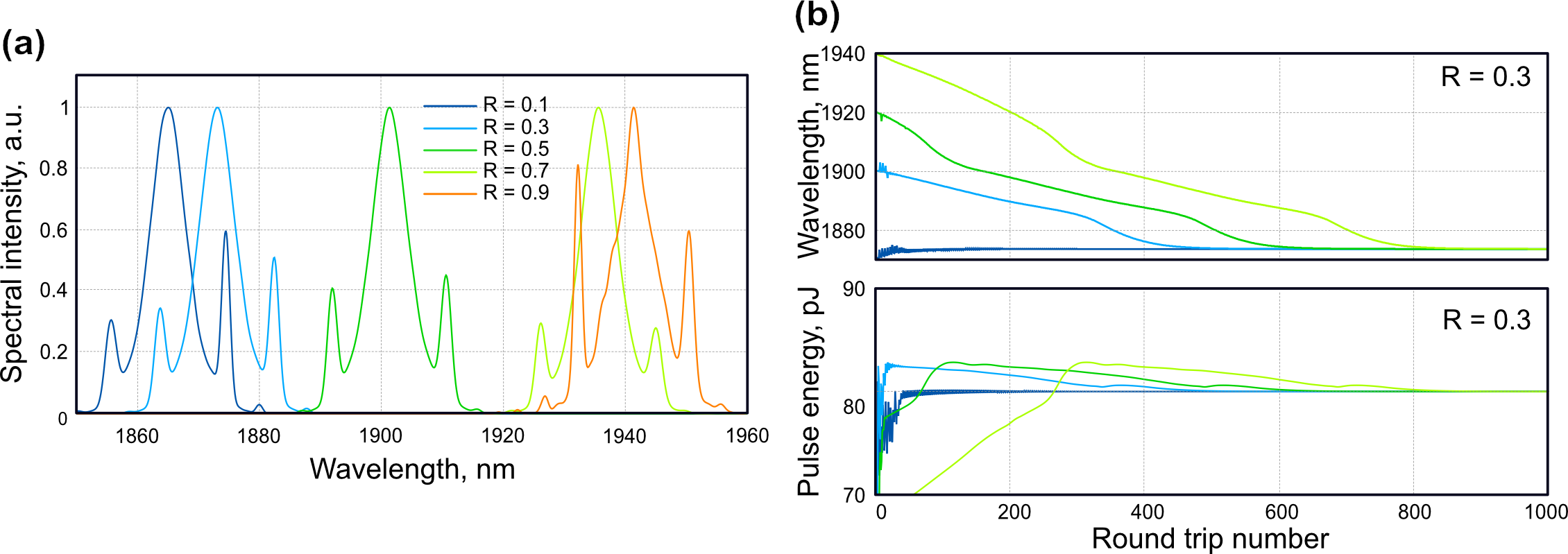}
    \caption{(a) Calculated output spectra corresponding to varying cavity feedback $R=0.1-0.9$. (b) Convergence to steady state regime of pulse generation with the increase of round trips number in simulations.}
    \label{fig:spectra_calc}
\end{figure} 

where $\sigma_{12}^p = 1.5630 \cdot 10^{-25}$ m$^2$ and $\sigma_{21}^p = 5.1005 \cdot 10^{-27}$ m$^2$ are the effective pump absorption and emission cross sections, $\sigma_{12}^s$ and $\sigma_{21}^s$ \cite{ixblueFibers} are the corresponding values for the signal wavelength $\lambda_i$, $N = 4.05938 \cdot 10^{15}$ m$^{-1}$  and $N_2$ -- are the total numbers of all and excited Tm-ions integrated over the fiber mode cross-section, $T = 250$ $\mu$s is the Tm-ion life time. $\rho_p$ and $\rho_s$ are normalized pump and signal power distributions through fiber cross-section $\rho_{p,s} = \Gamma_{p,s}/\pi a^2$,
where $a = 2.65$ $\mu$m - core radius of a single-mode fiber, $\Gamma_p (\Gamma_s)$ corresponds to the modal overlap factor between the pump (signal) mode and the ion distribution. $\Gamma_p = 1$ for core pumping, $\Gamma_s = 1-exp(-2a^2/w^2)$, $w$ -- 1/e electric field radius of the equivalent Gaussian spot \cite{Whitley:250058}.

The simulations start from the seed pulse at the first round trip and run until the steady state is reached. 

\begin{figure}[!]
    \centering
    \includegraphics[width=0.45\textwidth]{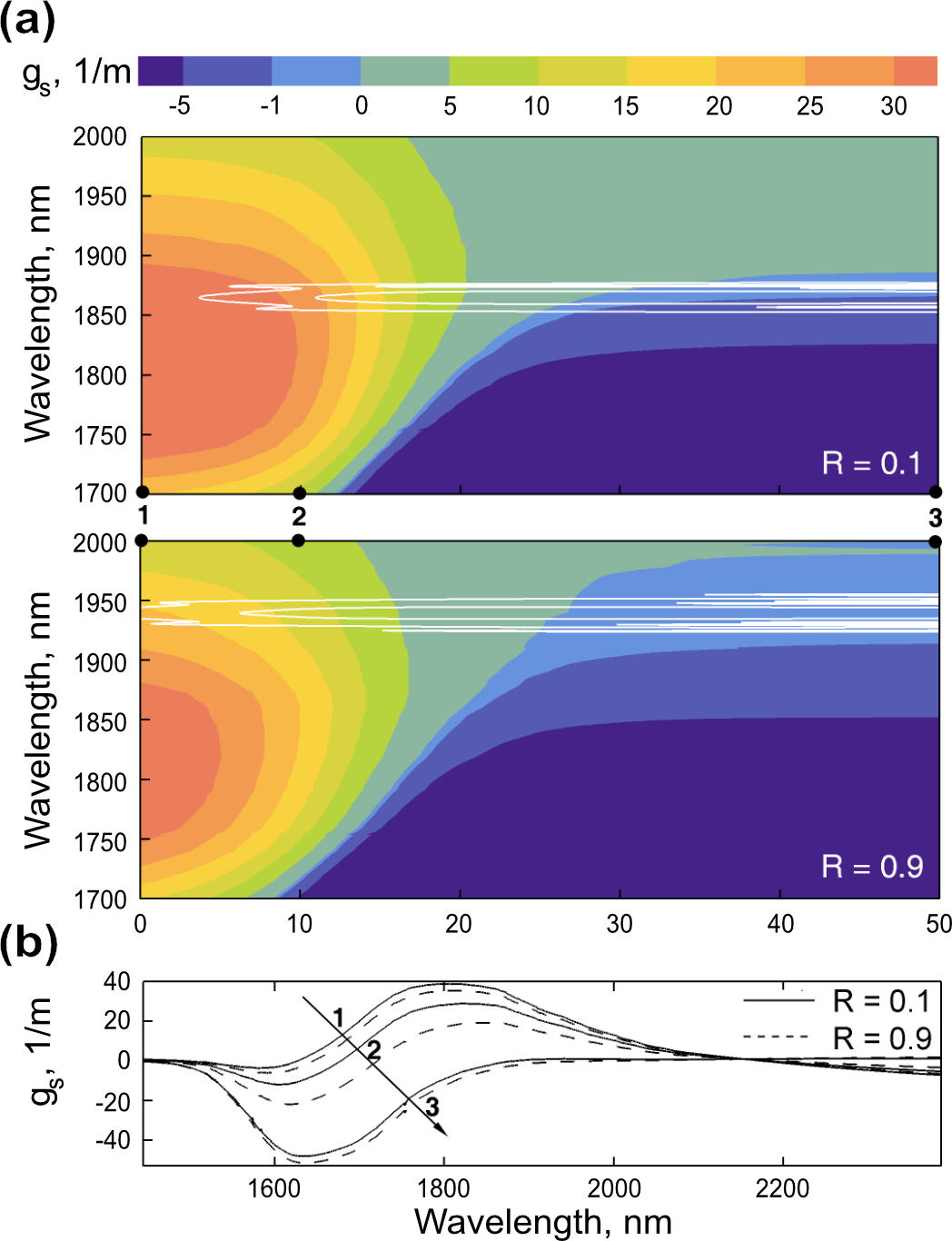}
    \caption{(a) Evolution of the gain profile along amplifier for R=0.1 (upper panel) and R=0.9 (bottom panel). Insets: White lines depict the pulse spectrum. (b) Gain spectra at different points (“1”, “2” and “3”) along the fibre at 10 and 90\% cavity feedback.}
    \label{fig:gain_evol}
\end{figure} 

\textbf{Figure~\ref{fig:spectra_calc}}\,a shows simulated output spectra at different gain excitation levels. The increase of cavity feedback leads to pulse central wavelength red-shift from 1865 to 1940~nm. The theoretically predicted wavelength tuneability is about 75~nm. The output spectra have pronounced Kelly sidebands characteristic to periodically amplified average solitons~\cite{kelly1992characteristic}. The pulse formation to the steady-state mode-locking regime was analysed at different parameters of the seed pulse. The results of numerical simulations show that the output pulse wavelength and energy in the stable regime do not depend on the wavelength and energy of the seed pulse at the first round trip. Figure~\ref{fig:spectra_calc}\,b shows the example of simulation with the feedback ratio of 30\%, where four initial pulses with different wavelengths (upper plots) and energies (bottom plots) converge to the same attractor with the increase of the cavity round trips number, demonstrating the uniqueness of the steady-state solution. This study justifies that the tuneability of the central operation wavelength in the filter-less laser cavity under investigation is governed solely by the gain spectrum dynamics.

To better understand pulse spectrum dynamics, we considered evolution of the gain $g_s(\lambda)$ along the active fiber corresponding to the cavity feedback values R=10\% and R=90\% (\textbf{Figure~\ref{fig:gain_evol}}). The pulse spectrum in steady-state is depicted by white lines in Figure~\ref{fig:gain_evol}\,a. Feedback increase to 90\% leads to higher pulse energy inside the cavity and faster gain saturation. The gain maximum shifts towards a higher wavelength and causes the shift of the pulse spectrum to find energy balance. It is worth noting that the wavelength of the intracavity pulse does not change along the active fibre in the steady-state regime, but is fixed and defined by the feedback value. Gain spectra measured at the points “1”, “2” and “3” along the fibre (Figure~\ref{fig:gain_evol}\,b) qualitatively agrees with the spectra shown in Figure~\ref{gain}, demonstrating the shift of the gain and losses maxima during the pulse amplification. Therefore, the results of numerical simulations confirm the idea of gain-controlled tuneability in a Tm-doped laser, as predicted in Figure~\ref{gain}.

\subsection{Experimental results}

\begin{figure}[b!]
\centering
  \includegraphics[width=0.95\textwidth]{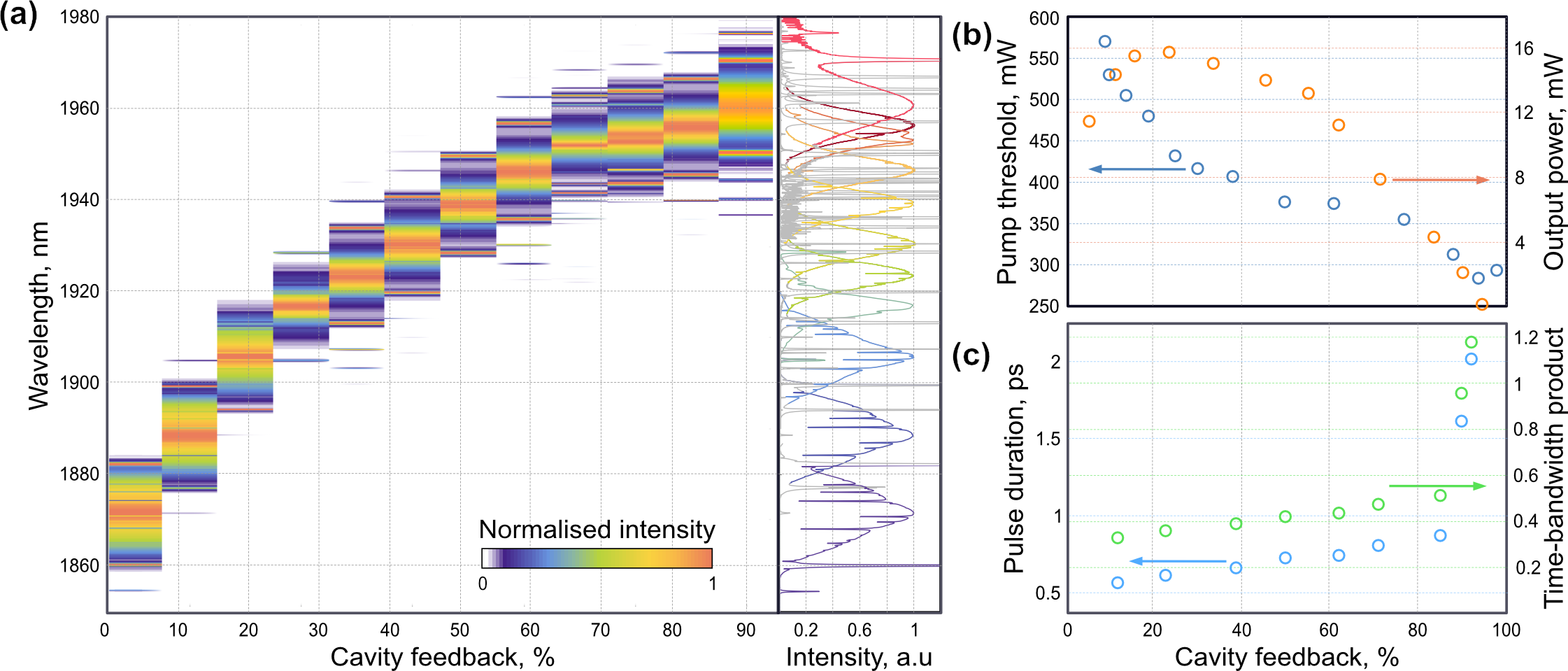}  
\caption{Tuneability of output pulse characteristics with the variable coupler. (a) Optical spectrum tuneability; (b) Generation threshold and efficiency; (c) Pulse duration variation.}
\label{fig:MeasuresVarCpl}
\end{figure}

We next examined the Tm-doped fibre laser with the variable output coupler experimentally. Here, mode-locked laser operation was observed within the entire range of the feedback tuneability from 8 to 93\%. With the pump power fixed at 730~mW and proper initial adjustment of intra-cavity polarisation controller, the continuous central wavelength tuneability was observed in the range spanning from 1873 and 1962~nm (\textbf{Figure~\ref{fig:MeasuresVarCpl}}\,a). As the numerical simulation predicted, the tuneability was highly reproducible with the feedback variation even after several switch-on and off cycles. \textbf{Supplementary video 1} demonstrates smooth wavelength tuneability with no pulse break up or instabilities. The variation of output average power is presented in Figure~\ref{fig:MeasuresVarCpl}\,c (blue scatters). This trend is in good agreement with generation efficiency dynamics predicted numerically using Rigrod model~\cite{Rigrod.1965s} (\textbf{Supplementary Figure S1}), which showed the highest efficiency at ~20\% cavity feedback. Figure~\ref{fig:MeasuresVarCpl}\,c demonstrates the pulse broadening from 550 to 860~fs with the feedback increase. While the spectral full width at half maximum ranges only from 7.4 to 7.7\,nm with the increase of the cavity feedback, the pulses evolve from transform-limited at ~12\% to slightly chirped. At high feedback values (over 85\%) and, therefore, high intracavity intensities, the time-bandwidth product rises to 1.172, indicating that nonlinear effects do not balance cavity dispersion.

\begin{figure}[b!]
\centering
  \includegraphics[width=0.95\textwidth]{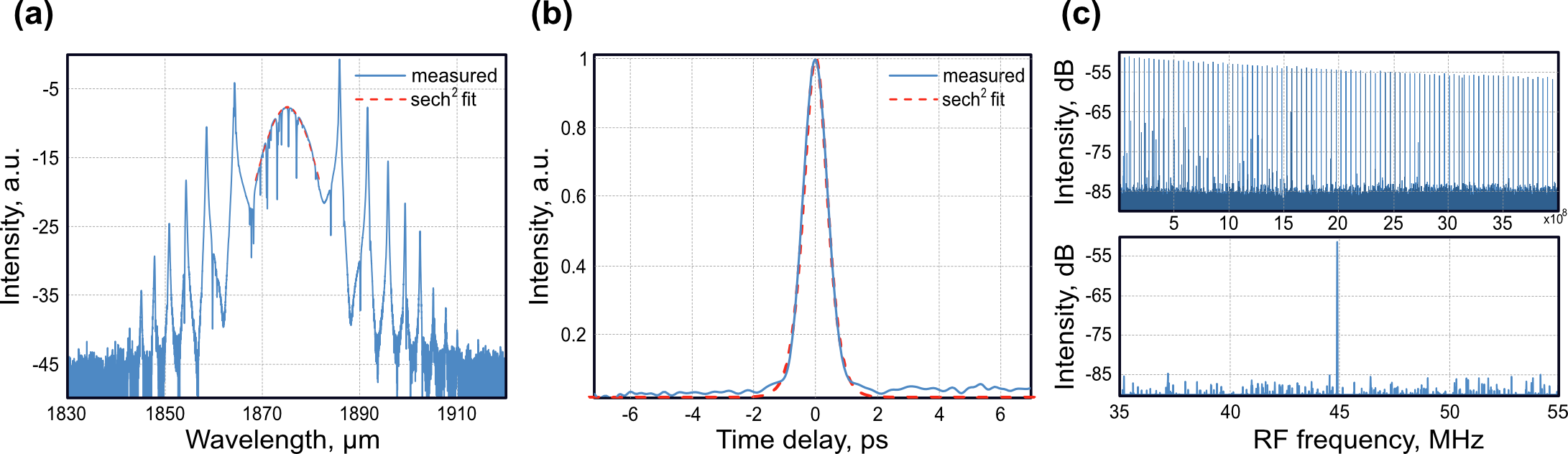} 
\caption{Pulse characteristics with (a) optical spectrum, (b)~autocorrelation trace, (c)~absolute frequency spectrum.}
\label{fig:highest}
\end{figure}

With the feedback increase, the pump power threshold for achieving self-mode-locking decreased from 538 to 289~mW (Figure\,\ref{fig:MeasuresVarCpl}\,b). At the same time, the upper threshold for stable single-pulse operation reduces with higher feedback due to gain saturation. With careful adjustment of the polarisation controller, maximum average output power of 68~mW could be obtained in a single-pulse generation regime with the 25\% feedback and 1240~mW pump power at 1889~nm central wavelength. \textbf{Figure~\ref{fig:highest}} demonstrates the output pulse parameters at the highest average power. The Figure\,\ref{fig:highest}\,b demonstrates autocorrelation trace at the highest average output power, generated in the cavity with 25\% feedback. The fundamental repetition rate was around 44~MHz with a decent 34~dB signal-to-noise ratio and good stability (Figure~\ref{fig:highest}\,c).

Furthermore, the generation regime can be adapted by using the feedback variation. \textbf{Supplementary Figure S2} shows the map of possible generation regimes with fixed polarisation controllers and with their adjustment ensuring single-pulse generation, when possible. As seen, generation can be switched from the fundamental solitons to soliton complexes, stable high-harmonics generation (up to fourth order), and chaotic behaviour.



\section{Discussion}

Our experimental and numerical studies of the self-mode-locked Tm-doped fibre laser with variable feedback demonstrated broadband central wavelength tuneability. Both studies have confirmed the variation of the gain excitation level through alteration of cavity feedback to be the primary mechanism of the tuneability of the ultrashort pulse spectrum. At the same time, initial seed conditions or intracavity nonlinear dynamics do not affect the central wavelength in a steady-state generation regime. We observed experimentally and justified theoretically, that the pump power increase, and, thus, corresponding increase of active fibre excitation, also does not lead to the significant alteration of the central wavelength range. 

However, despite the good qualitative agreement, the experiments showed a larger tuneability range than predicted in the numerical model. We assign this discrepancy to certain complexities that were bypassed or could not be included in this laser model. Particularly, the model considered standard silica glass matrix doped with Tm$^{3+}$ ions. Whereas in the experiment, Tm-doped fibre has a more complicated glass composition, including fluorine and phosphorous, which shifts Tm$^{3+}$ emission spectrum. Furthermore, the tuneability range may be enlarged by implementing a higher gain factor, for instance, by application of elongated Tm-doped fibre or fibre with higher doping concentration, provided there is an adequate pump power.

\begin{figure}[!]
\centering
  \includegraphics[width=0.45\textwidth]{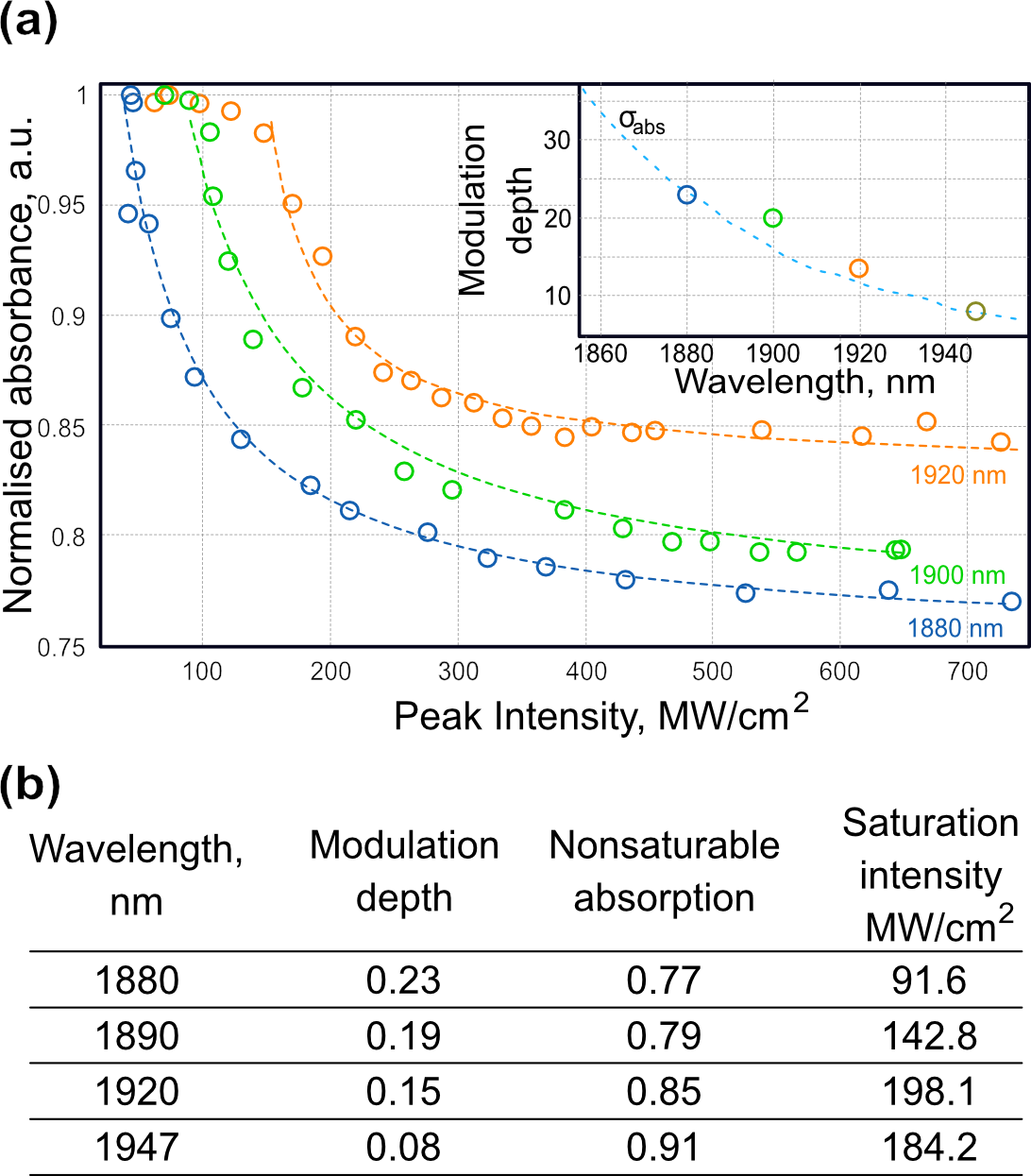}
  \caption{Normalised nonlinear absorption of 5\,cm-long section of the Tm-doped fibre at different wavelengths. (a) Measured values; Inset: correspondence of the absorption cross-section spectrum with the measured modulation depth. (b) Table of the deduced  parameters.}
  \label{fig:Tm-SA measurment}
\end{figure}

Another uncertainty in the model attributes to the nonlinear absorption properties of the used active fibre. The variation of stored energy in the laser cavity would obviously cause the alteration of the non-excited length of the active fibre, acting as a saturable absorber. Naturally, the length of the active fibre, as a distributed saturable absorber, determines its modulation depth and saturation intensity, the deep study on which we discuss in follow up paper. In addition, these parameters significantly depend on the operation wavelength.
To assess the influence of the wavelength tuneabiltiy on the nonlinear absorbance of the Tm-doped fibre used in the experiment,  we investigated the nonlinear behaviour of 5-cm long section at the wavelength range spanning from 1880 to 1947~mW. The measurements were achieved by launching 500-fs pulses at 44~MHz repetition rate into the fibre under test. {\bf Figure}~\ref{fig:Tm-SA measurment} shows normalised power-dependent absorption measurement of the Tm-doped fibre at 1880, 1900, and 1920~nm, which is in good agreement with the approximation according to the two-level energy model \cite{Sheik.1989h,silfvast2004laser}. The summary of the parameters of Tm-doped fibre as a saturable absorber at different wavelengths is presented in Figure~\ref{fig:Tm-SA measurment}\,b.

As shown in the inset in Figure\,\ref{fig:Tm-SA measurment}, the trend of the measured values of the saturable loss coincides well with the relative absorption slope of Tm$^{3+}$ ions. The saturation intensity depends on the upper state lifetime $\tau$ and the absorption cross-section $\sigma_{abs}$ expressed as $I_{sat} = h c (2 \lambda \tau \sigma_{abs})^{-1} $, such that the modulation depth is proportional to the absorption cross-section, taking Equation~\ref{alpha} into account~\cite{BergmannSchaefer}. While considering the variability of modulation depth and saturation intensity with the wavelength in the model would affect the generation regime through alteration of gain/loss balance via insertion losses. Yet, the saturable absorber model already includes dependence on launched intensity. The additional inclusion of the second variable of the wavelength would make the model significantly more complicated, which is outside the current study on laser tuneability.

\section{Conclusion}

The experiments and modelling investigations reported in the current work have extended the understanding of gain spectrum dynamics in ultrafast fibre lasers. The most substantial result is the justification of the pivotal role of the gain excitation level in controlling generation wavelength. Thus,  through alteration of cavity feedback in ultrafast Tm-doped fibre laser, we have recorded generation central wavelength within the range spanning 1873 to 1962~nm. The control of the excitation level of the gain through variation of out-coupled power proved to be more advantageous than the loss management, allowing to achieve of maximum 1~nJ output pulse energy with $\sim$915~mW pump power, resulting in 6.5\% optical conversion efficiency; or a differential efficiency of 22\% at most.

From an instrument development viewpoint, our results have provided a further example of the great versatility of ultrafast fibre lasers. Although the current studies were focused on Tm-doped fibres as the gain medium, the key underlying phenomenon of the suggested tuneability method is general and could be translated to other wavelengths, where the majority of fiberised laser components, including filters, are currently unavailable. In particular, this refers to the exploration of the Mid-IR wavelength range, where Dy and Er-doped fluoride-based fibres also offer exceptionally broad gain spectra.

\section{Methods}\label{ExpSec} 


\threesubsection{Instrumentation}\\
The measuring equipment is made up of a 10\,GHz extended InGaAs photodetector (ET-5000, EOT) next to a 25\,GS/s oscilloscope (DPO\,70604C, Tektronix) for electric characterisation. For optical characterisation an optical spectrum analyser with down to 50\,pm resolution and a 1.1--2.5\,\textmu{m} sprectral coverage is used (AQ6375, Yokogawa). With a frequency doubling autocorrelator (PulseCheck, APE), the time dependence has been examined. For quantifying the mean optical power, an integrating sphere type sensor with InGaAs photodiode and 1\,nW resolution is employed (S148C, Thorlabs).

\medskip
\textbf{Supporting Information} \par 
Supporting Information is available from the Wiley Online Library or from the author.

\medskip
\textbf{Acknowledgements} \par 
D.C.K. and M.C. acknowledge the support of the Deutsche Forschungsgemeinschaft (DFG – German Research Foundation)
under project CH2600/1-1. A.B. acknowledge the support of the Ministry of Science and Higher Education of the Russian Federation (Project No. FSUS-2021-0015).
\medskip

\textbf{References}\\

\bibliographystyle{MSP}
\bibliography{2Bibliography}

\end{document}